\documentclass[pra,twocolumn,unsortedaddress,superscriptaddress]{revtex4-1}
\usepackage{graphicx,amssymb,amsmath,latexsym,epsfig}
\usepackage{color}
\newcommand{\e}[1]{\mbox{e}^{#1}}

\newcommand\rf[1]{Eq. (\ref{#1})}

\newcommand{\cop}[2]{\hat{#1}^\dagger_{#2}}
\newcommand{\aop}[2]{\hat{#1}_{#2}}

\newcommand{\p}{\mathbf{p}}
\newcommand{\x}{\mathbf{x}}

\newcommand{\rr}{\mathbf{r}}
\newcommand{\y}{\mathbf{y}}

\newcommand{\kk}{\mathbf{k}}
\newcommand{\grad}{\vec{\nabla}}
\newcommand{\xplus}{X_+}
\newcommand{\xmin}{X_-}
\newcommand{\vxplus}{\mathbf{X}_+}
\newcommand{\vxmin}{\mathbf{X}_-}
\newcommand{\cgrad}{\overleftrightarrow{\nabla}}
\newcommand{\cpartial}{\overleftrightarrow{\partial}}
\newcommand{\re}{\text{Re}}
\newcommand{\im}{\text{Im}}

\begin{document}
\title{``Charged" phonons in an external field:  a QED analog with Bose-Einstein condensates}

\author{Shay Leizerovitch}
\affiliation{School of Physics and Astronomy, Raymond and Beverly Sackler Faculty of Exact Sciences, Tel Aviv University, Tel-Aviv 69978, Israel.}
\author{Benni Reznik}
\affiliation{School of Physics and Astronomy, Raymond and Beverly Sackler Faculty of Exact Sciences, Tel Aviv University, Tel-Aviv 69978, Israel.}

\vskip.3in

\begin{abstract}
We propose a method for using ultracold atomic Bose-Einstein condensates, to form an analog model of a relativistic massive field that carries ``charge" and interacts with an external non-dynamical gauge field.
Such a ``scalar  QED" analog model, may be useful for simulating various effects of  QFT involving charged particles, such as the Schwinger pair-creation of ״charged״ phonons in a constant external field, and vacuum instability.
\end{abstract}
\maketitle
\section{Introduction}
The origins of the idea, that nature manifests close analogies between phenomena at very different scales, can be traced back to the dawn of science.  ``Analogous arguments" have been for example used to motivate the first atomistic theory of nature – as beautifully
described by Lucretius \cite{lucretius}.  
In contemporary physics it turns out that the idea of nature analogies is reflected more concisely in the mathematical resemblance of phenomena at very different length and time scales; from the High Energy scales of particle physics which are described by High-Energy Physics  (HEP), to non-relativistic atomic many-body systems, and further up to the large
cosmological scales and black-holes in General Relativity (GR)\cite{Wilczek1998,Unruh2007book,volovik2009}.
 
One research field, that has particularly benefited by employing analogies, has emerged from Hawking's discovery that quantum mechanical effects cause black-holes to emit  radiation and gradually evaporate \cite{Hawkingnature,Hawking1975}. In the absence of a complete theory of quantum gravity the black-hole evaporation effect, is unfortunately still only partially understood even at the fundamental level \cite{Hawking1976}. 
A  ``fluid-model" of a black-hole proposed by Unruh \cite{Unruh}, provided an analog framework  for studying Hawking radiation, and inspired theoretical \cite{Jacobson1991,Jacobson1993,Unruh1996,Reznik1997,Corley1998,Visser} and experimental ideas \cite{GarayPRL,GarayPRA,Reznik2000,Leonhardt,Fedichev,Philbin,BenniH,Horstmann2010,Horstmann2011,Elazar}.
Remarkably, recent experiments, notably with ultracold atomic condensates that mimic
a black-hole, provided the first confirmations of the Hawking effect
\cite{Steinhauer2010,Weinfurtner2011,Steinhauer2014,Steinhauer2015}.
 
While atomic condensates turned helpful in gravitational scenarios,
they have been somewhat more restrictive in connection with HEP problems, and so far mostly used to mimic free relativistic fields.
The ``building-blocks" of the Standard Model  of high-energy physics, are gauge field theories  (abelian and non-abelian). They are essential for understanding effects such as quark confinement in QCD, and other effects. Recently it has been proposed that such models, when formulated on a lattice, can be naturally simulated using discrete set-ups involving ultracold atoms traped in optical lattices \cite{Erez2015,Cirac2010,ErezPrl,Wiese,Szpak,Zohar2016}.
 
In this article we shall suggest however, that the analogy between atomic condensates and quantum field theory can be extended and mimic charged massive relativistic fields under the influence of an external electromagnetic gauge field. 
We shall show that a QED analog model can be constructed using several Bose-Einstein condensates, with suitably chosen interactions. 

The article proceeds as follows:
In section \ref{sec:gauge_theory} we discuss the basic features of quantum field theory coupled to external gauge field.
We introduce the dynamical degrees of freedom of the theory, complex scalar fields, and introduce their interaction with the gauge field.
In section \ref{sec:massive_phonons}, we show following \cite{Visser}, (using the Bogoliubov Hamiltonian) that coupled two-mode condensates give rise to a massive phonon mode that is analogous to a massive Klein-Gordon field. 
In section \ref{sec:lorentz_invariance}, we construct a real scalar field and its conjugate momenta using massive excitations. We then discuss the emergence of Lorentz invariance and locality.
In section \ref{sec:charge} we further extend the condensate system and build an analog complex scalar field.
We identify the charge operator, associated with global phase invariance, and identify positive and negative phonon charge carriers.
Finally, in section \ref{sec:intHam} we construct the interaction between a charged field and an external gauge field.

\section{ Local gauge invariance: scalar QED}\label{sec:gauge_theory}

We begin by briefly discussing  the basic key features of quantum field theory that involve matter interacting with gauge fields.
Generally speaking, we consider the simplest example of a relativistic quantum field theory for dynamical charged matter that is coupled to a gauge field, that manifests the following two fundamental key features: invariance under space-time Lorentz transformations, and local gauge symmetry.
As opposed to the global nature of the ``external" space-time Lorentz transformations, the gauge symmetry represents the invariance of the theory under local transformations of ``internal" degrees of freedom.
The first property is essential to any relativistic field theory.
The second property, models with dynamically locally conserved charge.
It is well known from Noether's theorem, that {\it local} (as opposed to global) gauge invariance give rise
to a dynamically conserved quantity, which in this case physically corresponds to a charge that is carried by the matter field.

We shall consider a theory that has a $U(1)$ gauge symmetry.
Matter will be represented by a complex scalar field $\Phi(x)$,  and the external gauge field by $A_\mu(x)$.  Such a theory should then remain invariant under the transformation
$$
\Phi(x)  \to e^{i\theta(x)} \Phi(x), \ \ \  \Phi^\dagger(x)  \to e^{-i\theta(x)} \Phi^\dagger(x)
$$
and
$$
A_\mu(x) \to A_\mu(x) - \frac{\partial \theta(x)}{\partial x^\mu}
$$
where $\theta(x)$ is an arbitrary real function. 

A simple example of such a theory can be described by the Lagrangian density 
\begin{equation}\label{theory_lagrangian}
\mathcal{L}= \left[ D_\mu \Phi(x)\right]^\dagger   D^\mu \Phi(x) - m^2 \Phi(x)^\dagger \Phi(x) 
\end{equation}
where $D_\mu = \partial_\mu + ieA_\mu$. It can  be readily seen that indeed the theory respects both, space-time and internal gauge symmetries. 
In particular it gives rise to a conserved (abelian) charge   $Q= \int J_0 dx$, 
where $J_\mu =i(\Phi^\dagger D_\mu\Phi -\text{h.c})$

We shall establish an analogy between the field theory described by Eq. \ref{theory_lagrangian}, and a system involving ultracold non-relativistic atomic condensates. Since the later 
is usually described by a non-relativistic Hamiltonian, 
it will be helpful to recast our QFT model in the Hamiltonian framework, as well. 

The corresponding Hamiltonian density $\mathcal{H}$, can be expressed in terms of
the field $\Phi$ and the canonically conjugate momentum $\Pi(x)$, obeying a commutation relation $ \left[\Phi(\x),\Pi(\y)\right]=i\delta (\x-\y)$.

We have  $\mathcal{H}= \mathcal{H}_0+\mathcal{U}_E+\mathcal{U}_B$, where $\mathcal{H}_0$ is 
the free non-interacting part 
\begin{equation}\label{relFreeHam}
\mathcal{H}_{0} = \Pi^{\dagger}\Pi+(\grad\Phi)^{\dagger}(\grad\Phi)+m^{2}\Phi^\dagger\Phi~,
\end{equation}
the term $\mathcal{U}_E$ describes the ``electric" interaction with an external scalar field $A_0(x)$,  
\begin{equation}\label{relIntHam}
\mathcal{U}_E = ieA^{0}\left(\Pi^{\dagger}\Phi^{\dagger}-\Phi\Pi\right)+e^{2}A_0 A^{0}\Phi^\dagger\Phi~,
\end{equation}
and $\mathcal{U}_B$ ``magnetic" interaction with an external vector field $\vec A(x)$,
\begin{equation}\label{vectorrelIntHam}
\mathcal{U}_B = ieA^{i}\Phi^{\dagger}\cpartial_{i}\Phi+e^{2}A_i A^{i}\Phi^\dagger\Phi~.
\end{equation}
Since we consider a particular case of static external fields, the terms referred above to scalar and vector interactions, indeed correspond to electric and magnetic interactions. Furthermore, Eqs. (3,4) have simple physical meaning.   
The first term on the r.h.s of \eqref{relIntHam} (that henceforth will be denoted as $\mathcal{U}_E^{(1)}$) describes the interaction of the charged field with the electric potential, $A^0 J_0$
where
\begin{equation}
J_0=ie\left(\Pi^{\dagger}(x)\Phi^{\dagger}(x)-\Pi(x)\Phi(x)\right)~.
\end{equation}
The second term on the r.h.s of \eqref{relIntHam} (henceforth denoted as $\mathcal{U}_E^{(2)}$) is quadratic in $A^0$.  In strong electric fields, it gives rise to quantum relativistic effects, such as pair-creation, and instability of the vacuum.
The first term on the r.h.s. of \eqref{vectorrelIntHam} (henceforth denoted as $\mathcal{U}_B^{(1)}$) is the well known $\vec A\cdot \vec J$ interaction. Finally, the last term, $\propto \vec A^2$ (henceforth denoted as $\mathcal{U}_B^{(2)}$), becomes important at strong magnetic fields. For example,  in the non-relativistic case,  it produces the Landau energy-level structure in two-dimensional electronic systems.

In the following, it will turn useful to recast the above Hamiltonian in terms of real scalar fields. 
It can be readily seen that the free Hamiltonian part \eqref{relFreeHam}, can be re-expressed in terms of two scalar fields,  $\phi_i$  $i=1,2$, and their  conjugate momenta.  The fields are given by real and the imaginary parts of the original complex field:  $\phi_1 = \sqrt2 \re{\Phi}$, $\phi_2= \sqrt2 \im{\Phi}$. Similarly, their 
conjugate fields, are obtained from $\Pi(x)$.
The free Hamiltonian in this representation, is then given by the linear combination
$\mathcal{H}_0= \mathcal{H}_0(\phi_1,\pi_1)+\mathcal{H}_0(\phi_2,\pi_2)$. 
We should notice however, that in this bi-field representation, the electric and magnetic interactions (\ref{relIntHam},\ref{vectorrelIntHam}) , contain 
 quadratic mixed terms, $\propto \phi_i\phi_j$.

\section{massive phonons in a BEC} \label{sec:massive_phonons}
 
The Bogoliubov spectrum of phonon excitations in a BEC is linear at small momenta, and hence analogous to a single massless scalar field. 
An elegant method that gives rise to a sepcetrum of massive relativistic particles, that is to ``massive phonons" in a BEC, has been suggested \cite{Visser}.
The scheme starts with a two-mode condensate system, and introduces a Raman coupling between the condensates. The resulting normal modes, of the uncoupled condensates are deformed by the coupling;  one of the two dispersion branches remains massless, and the second acquires effective mass.
In the  hydrodynamical approximation, the latter corresponds to a relativistic massive particle with a Klein-Gordon energy-momentum-relation $E=\sqrt{p^2+ m^2}$.

In the following we consider a Hamiltonian density of two condensates 
\begin{equation}\label{GPhamiltonian}
\mathcal{H}= \mathcal{H}_{GP}+\mathcal{H}_L~,
\end{equation}
where 
\begin{equation}
\mathcal{H}_{GP} = \sum_{i,j=1}^2 \frac{\delta_{ij}}{2m}|\nabla{\psi}_{i}(\x)|^2  + U_{ij} |{\psi}^{\dagger}_{i}(\x){\psi}_{j}(\x)|^2
\end{equation}
and 
\begin{equation}\label{laser_int}
\mathcal{H}_L= \Omega_{M}(\psi_{2}^\dagger (\x)\psi_{1}(\x)+\mbox{h.c.}) 
\end{equation}
are the free Gross-Pitaevskii Hamiltonian density and laser (or micro-wave) induced interaction, respectively. $m$ is the mass of the condensate atoms,  $U_{ij}$ are the scattering coefficients, and $\Omega_M$ is the Rabi frequency.
The total Hamiltonian is  
 ${H} =   \int  d^3\x \mathcal{H}$.

\begin{figure}[t]
                \includegraphics[width=0.4\textwidth]{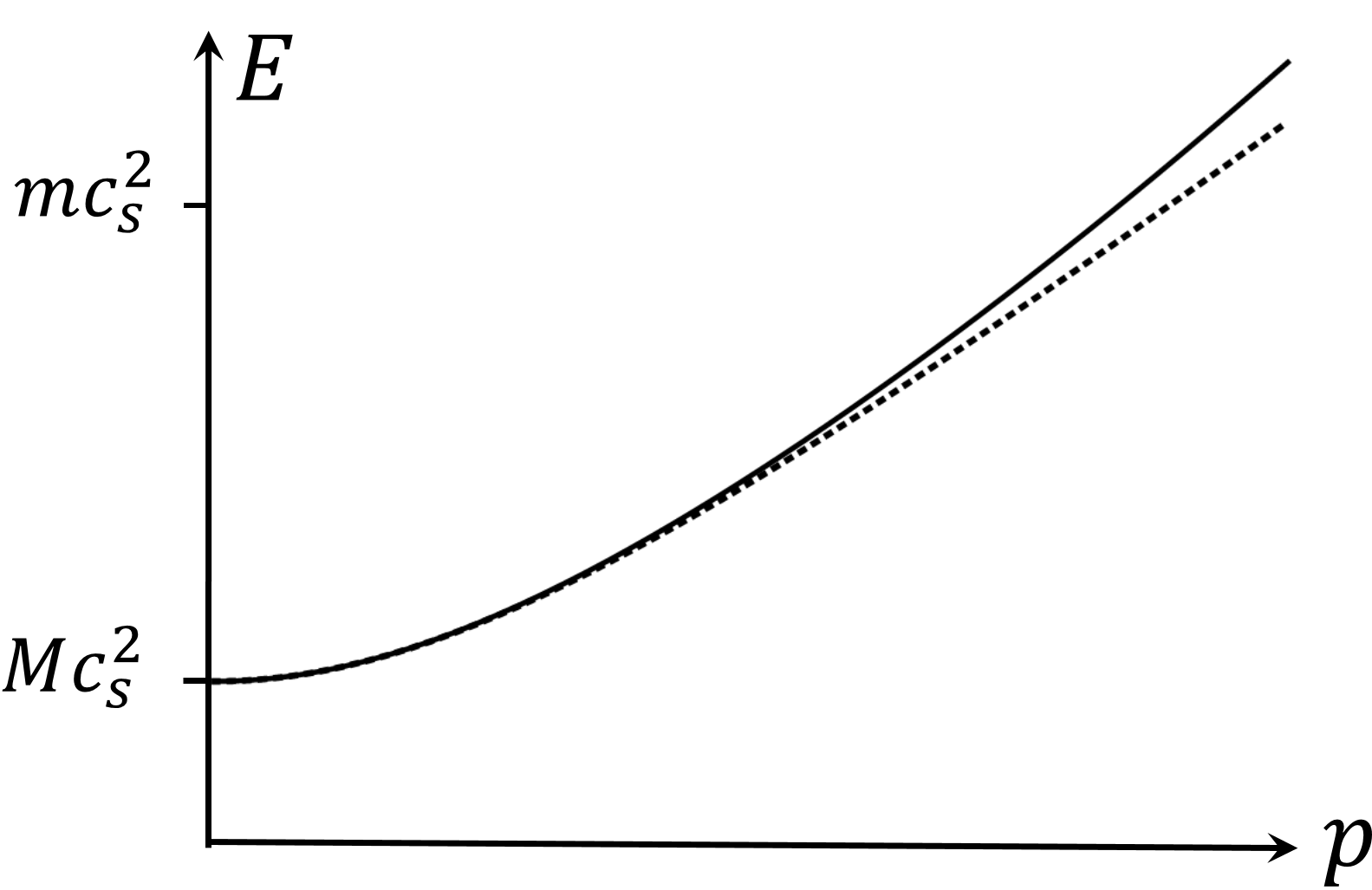}
                \caption{Gapped dispersion relation. The dotted line denotes an ideal Klein-Gordon energy. The solid line corresponds to the mode $J=M$. 
Beyond the cut-off energy scale, $E_{\text{cut-off}} \sim mc_s^2$, the two energies begin to deviate.}
                \label{fig:dispersion}
\end{figure}

By expending  the condensates' fields, $\psi_i$,  around their mean field $\bar{\psi}_{i}$, as ${\psi}_i(\x)=\bar{\psi}_{i}+\delta{\psi}_i(\x)$, where  $\delta{\psi}_i(\x)$ are small enough perturbations, and furthermore  assuming that the condensates are uniform and that  $\bar\psi_{1}=\bar\psi_{2}=\sqrt{n}$,
we can approximate ${H}$ by a quadratic Hamiltonian  in terms of $\delta\psi_{i}(\x)$.
The latter can be diagonalized by expanding
\begin{equation}\label{expansion}
\delta\psi_i(\x) = \int d^3\p \,a_{i,\p}e^{i\p\x} ~,
\end{equation}
and then performing a Bogoliubov transformation (see appendix \ref{app:diagonalization}).
For the following it will be convenient to redefine the integration on the momentum space as
\begin{equation}
\int d^d\p\equiv \int \frac{d^{d}\p}{(2\pi)^{d/2}}~.
\end{equation}
The Bogoliubov transformation leads to:
\begin{equation}\label{bog_trans}
a_{i,\p} = \frac{1}{\sqrt{2}}\sum_{J=(0,M)} \bigl( u_{i,J}(\p) b_{J,\p} + v_{i,J}(-\p) b^\dagger_{J,-\p} \bigr)~.
\end{equation}
The first normal mode,  $J=0$, turns out gapless.
The second normal mode, $J=M$, is gapped with $E=Mc_s^2$ at $\p=0$, as illustrated in Fig. \ref{fig:dispersion}. 
 
The Bogoliubov amplitudes are given by
\begin{equation}\label{qpamplitude_ML}
u_{i,0}(\p),v_{i,0}(-\p)=\pm\sqrt{\frac{ \left(\text{mc}_{s0}^2+\epsilon _p\right)}{2E_{0}(p)}\pm \frac{1}{2}}~,
\end{equation}
and
\begin{equation}\label{qpamplitude}
u_{i,M}(\p),v_{i,M}(-\p)=\pm(-1)^{i}\sqrt{\frac{ \left(\text{mc}_{sM}^2+\epsilon _p\right)}{2E_{M}(p)}\pm \frac{1}{2}}~,
\end{equation}
where $\epsilon_p=p^{2}/2m$.
The complete expressions for the massless and  massive dispersion branches are given in appendix \ref{app:diagonalization} (They may be of much interest, for example close to sonic black-hole horizons which mixes low and high frequencies \cite{Jacobson1991,Jacobson1993,Unruh1996,Corley1998,Reznik2000}).
For small energies, the gapped mode agrees with the dispersion relation 
\begin{eqnarray}\label{relappen2}
E(p)=\sqrt{M^2c_{s}^4+c_{s}^2p^2}~,
\end{eqnarray}
of a massive relativistic particle of rest mass energy
$Mc_s^2$,
\begin {equation}\label{mass_relation}
\frac {M^2}{m^2}= \frac {-4\Omega_M(n(U-U')-\Omega_{M})}{(n(U-U')- 2 \Omega_M)^{2}} ~,
\end{equation}
where the upper limit on the speed of (massive) sound waves, $c_{sM}\equiv c_s$, is define by
\begin{equation}
m c_s^2 = {n(U-U')- 2\Omega_M}~.
\end{equation}
We observe that the effective mass is imaginary, and may lead to dynamical instability of the system \cite{pitaevskii2003}. In order to avoid it, we set $\Omega_M$ to be negative.
Since the expression for $E(p)$ above, holds only below the cutoff energy scale,  $E(p)\ll mc^2_s$, it also follows that the rest mass $M$, of the massive phonon is smaller then the mass $m$ of the particles that form the condensates: $M\ll m$.
If we demand that the dispersion relation remains valid in the ultra relativistic regime,
 $E(p)=\sqrt{p^2+M^2} \approx c_s p $, we must further require $Mc_s \ll p \ll m c_s  $.
Using \eqref{mass_relation}, we can see that this bounds the Rabi frequency of coupling laser 
\begin{equation}
(2mL^{2})^{-1}<\Omega_M\lesssim \varepsilon n(U-U^{\prime})/2~,
\end{equation}
where $\varepsilon$ is the ratio of the masses $M/m$, and $L$ is the typical length of the system. Therefore, setting the effective mass to be one order of magnitude less than the atomic mass, leads to the bound $|\Omega_M|\lesssim 0.05 n(U-U^{\prime})$.

Having identified the massive and massless normal modes we return to discuss the spatial structure of the field operators $\delta{\psi}_i(\x)$.
Using Eqs. (\ref{expansion},\ref{bog_trans}), we can rewrite  $\delta{\psi}_i(\x)$ as
\begin{equation}\label{eigenmodes}
\delta{\psi}_i(\x)=\frac{1}{\sqrt{2}}\left(\varphi_0(\x)+(-1)^i\varphi_M(\x)\right)
\end{equation}
where $\varphi_M(\x)$ and $\varphi_0(\x)$, are complex massive and massless condensate field operators respectively. They satisfy the commutation relations
\begin{equation}
 [\varphi_J(\x),\varphi^\dagger_{J'}(\x^{\prime})] = \delta_{J,J'}\delta(\x-\x^{\prime})~.
\end{equation}
We see that the  massive spatial field is given by the ``relative" field: $\varphi_M = (\delta{\psi}_2(\x)-\delta{\psi}_1(\x))/\sqrt{2}$, while the massless field correspond to the ``collective" combination: $\varphi_0 = (\delta{\psi}_1(\x)+ \delta{\psi}_2(\x))/\sqrt{2}$.
In the following we omit the index $J$,  and use $\varphi(\x)$, to denote a massive field.
 
The complex field $\varphi(\x)$, can now be used to construct a single massive scalar field,  $\phi(\x)$,  and a corresponding conjugate momentum, $\pi(\x)$,  
as follows: 
\begin{eqnarray}
\phi(\x)&=& \frac{i}{\sqrt{2}}(\varphi(\x)-  \varphi^\dagger(\x) )
\label{map_relfield}\\
\pi(\x)&=& \frac{1}{\sqrt{2}}(\varphi(\x)+ \varphi^\dagger (\x))~. \label{map_conmomenta}
\end{eqnarray}
The latter satisfy the canonical commutation relation
\begin{equation}\label{can_cr}
[\phi(\x), \pi(\x')] = i\delta(\x-\x')~.
\end{equation}
The above conjugate real fields can be expressed in terms of the Bogoliubov  modes as
\begin{eqnarray}
\!\!\!\!\!\!\phi(\x)&=& \frac{i}{\sqrt{2}}\int\!\! d\p\bigl(u(\p)-v(\p)\bigr)\left(\aop{b}{\p}\e{i \p\x}-\cop{b}{\p}\e{-i\p\x}\right)\label{relfield_bog_amps}\\
\!\!\!\!\!\!\pi(\x)&=&\frac{1}{\sqrt{2}}\int\!\! d\p\bigl(u(\p) +v(\p)\bigr)\left(\aop{b}{\p}\e{i \p\x}+\cop{b}{\p}\e{-i\p\x}\right)\! .\label{conmomenta_bog_amps}
\end{eqnarray}
When considering the relevant regime,  $E(\p)\ll E_{\rm cutoff} = mc_s^2$, we can use the relation
\begin{equation}\label{prefactors}
(u(\p)+v(\p))^2=(u(\p)-(\p))^{-2}\approx\frac{E(p)}{mc_s^{2}}~,
\end{equation}
and readily obtain 
\begin{eqnarray}\label{fieldsmap}
\phi(\x)&=& \frac{i}{\sqrt{2}}\int d\p \sqrt{\frac{mc^2_s}{E(p)}} \biggl(\aop{b}{\p}\e{i \p\x}-\cop{b}{\p}\e{-i\p\x}\biggr)\label{relfield}\\
\pi(\x)&=&\frac{1}{\sqrt{2}}\int d\p
\sqrt{\frac{E(p)}{mc^2_s}}
\biggl(\aop{b}{\p}\e{i \p\x}+\cop{b}{\p}\e{-i\p\x}\biggr)~.\label{conmomenta}
\end{eqnarray}
This coincides with the normal mode expansion for a scalar field, up to a trivial re-definition: $\aop{b}{\p}\rightarrow i\aop{b}{\p}$.
Since the $E(p)$ above is given by the Klein-Gordon dispersion relation, it is straightforward to verify that the fields satisfy the Klein-Gordon equation.

\section{Emergence of Lorentz invariance}\label{sec:lorentz_invariance}

Although the previous sections shows that the KG Hamiltonian is obtained in the low energy limit, it is constructive to derive it starting from more general assumptions.
Let us consider the excitation field $\varphi(\x)$ expansion in terms of the creation operators $\aop{b}{\p}$. 
The  scalar field and conjugate momenta satisfy 
\begin{equation}\label{operators_to_field}
\frac{1}{\sqrt{2}}(\pi(\x)-i\phi(\x))={\varphi}(\x)=\int d\p (u_p\aop{b}{\p}e^{i\p\x}+v_p\cop{b}{\p}e^{-i\p\x})~.
\end{equation}
We invert the above equation and express the creation operators  $\aop{b}{\p}$ in terms of the field and its conjugate momentum. Then by substituting in the Hamiltonian
${H}=\int d\p E(p)\cop{b}{\p}\aop{b}{\p}$, we obtain
\begin{eqnarray}\label{nonlocal_Ham}
{H}=\frac{1}{2}\int d\p E(p)\int d\x &d\y & e^{i\p(\x-\y)}[(u-v)^{2}\pi(\x)\pi(\y)\nonumber\\
&&+(u+v)^{2}\phi(\x)\phi(\y)]~.
\end{eqnarray}
Due to the dependence of the latter on the momentum, the resulting Hamiltonian has a non-local structure. It does not correspond to a local and Lorentz invariant QFT.
However, when considering the low energy limit \eqref{prefactors},  the Hamiltonian reduces to
\begin{eqnarray}
{H}&\approx &\frac{1}{2}\int d\p\int d\x d\y e^{i\p(\x-\y)}[\pi(\x)\pi(\y)+E^{2}(p)\phi(\x)\phi(\y)]\nonumber\\
&=&\int d\x[\frac{1}{2}\pi^{2}(\x)+\frac{1}{2}\left|\nabla\phi(\x)\right|^{2}+\frac{1}{2}M^{2}\phi^{2}(\x)]~.
\end{eqnarray}
Hence, locality, and Lorentz invariance, both emerge in low energy limit. The effective low energy theory corresponds, as expected, to a massive free QFT.

\section{``Charged" Excitations}\label{sec:charge}

Charged particles can be described by a single complex field that satisfies the Klein Gordon equation. 
Alternatively, we can represent the charged field, in terms of a pair of massive real scalar fields, of the same mass.
In the case of a free charged field, the fields remain decoupled and hence can be easily constructed using the method described section \ref{sec:massive_phonons}. 

We shall therefore consider a four level condensate, $\psi_{i}$ with $i=1,...,4$. The Raman interaction term \eqref{laser_int} involves two pairs of (laser) coupled condensates: 
 ${\cal{H}}_L =(\psi_{2}^\dagger (\x)\psi_{1}+\psi_{4}^\dagger (\x)\psi_{3} (\x)+\mbox{h.c.}) $.
Clearly the relative complex fields, $\delta\psi_{2k}
- \delta\psi_{2k-1}$, $k=(1,2)$, correspond to two massive complex condensate field operators, and following the same construction as in Eqs. (\ref{map_relfield},\ref{map_conmomenta}), 
we obtain a pair of real massive scalar fields that satisfies the commutation relations
\begin{equation}\label{complex_can_cr}
[\phi_i(\x), \pi_j(\x')] = i\delta_{ij}\delta(\x-\x')
\end{equation}
with $i,j = (1,2)$.

Finally we can build out of the latter a single complex field $\Phi(\x)$, and conjugate momentum $\Pi(\x)$ as
\begin{equation}
\Phi(\x)= \frac{\phi_1(\x)+i \phi_2(\x)}{\sqrt2}
\end{equation}
\begin{equation}
\Pi(\x)= \frac{\pi_1(\x)-i \pi_2(\x)}{\sqrt2}
\end{equation}
such that
\begin{equation}
\left[\Phi(\x),\Pi(\x^{\prime})\right]=i\delta(\x-\x^{\prime})~.
\end{equation}

The complex field can be decomposed in terms of orthogonal field operators $\aop{c}{p}$ and $\aop{d}{p}$ as
\begin{equation}
\Phi(\x)=\int \frac{d\p}{\sqrt{2E}}(\aop{c}{\p}e^{i\p\x}-\cop{d}{\p}e^{-i\p\x})~,
\end{equation}
where 
\begin{equation}
[\aop{c}{\p},\cop{c}{\p^{\prime}}]=[\aop{d}{\p},\cop{d}{\p^{\prime}}]=\delta_{\p,\p^{\prime}}~.
\end{equation}
Comparing equations (\ref{complex_can_cr}) and (\ref{relfield}-\ref{conmomenta}), we find  that in terms of condensate excitations', the charged excitations are created by  
\begin{equation}
\begin{cases}
\aop{c}{\p}=\frac{1}{\sqrt{2}}(i\aop{b}{1,\p}-\aop{b}{2,\p})\\
\aop{d}{\p}=\frac{1}{\sqrt{2}}(-i\aop{b}{1,\p}-\aop{b}{2,\p})
\end{cases}~.
\end{equation}

So far, our system is invariant only under global U(1) (phase) transformations. In the following sections, once the external gauge field will be included, the $U(1)$ symmetry will become local, and through Noether's theorem, the charge will be guaranteed to be  locally conserved. 
We can however, identify already the operator that corresponding to the global charge,
\begin{equation}
\mathcal{Q}= ie\int d\x\left(\Pi^{\dagger}(\x)\Phi^{\dagger}(\x)-\Phi(\x)\Pi(\x)\right)~,
\end{equation}
where $e$, the unit charge, has no physical significance in the absence of interaction.
Substitution of the complex field and its conjugate momentum in terms of $\aop{c}{\p}$ and $\aop{d}{\p}$ yields
\begin{equation}
\mathcal{Q}= ie\int d\p\left(\cop{c}{\p}\aop{c}{\p}-\cop{d}{\p}\aop{d}{\p}\right)~.
\end{equation}
We observe that excitations of type $c$ and type $d$, carry opposite ``charges":  type $c$ carries positive charge, and  type $d$ carry negative charge. They can be regarded as ``anti-particles" of each other.
In terms of the condensates' degrees of freedom this gives
\begin{equation}\label{charge_operator}
\mathcal{Q}= ie\int d\p(\cop{b}{2,\p}\aop{b}{1,\p}-\cop{b}{1,\p}\aop{b}{2,\p})~.
\end{equation}
Charge conservation, can be seen here as due to the  commutativity of the latter with the number operator, and thus with the free Hamiltonian
\begin{equation}
\left[\mathcal{Q},H\right]=0
\end{equation}
(For a free field, it is conserved for each $p$, separately). 

Finally we note, that due to gauge invariance, the Hilbert space of our system can be written as direct sum of different (total) charge sectors
\begin{equation}
\mathcal{H}=\bigoplus_Q\mathcal{H}_{Q}~.
\end{equation}
Charge conservation forbids transitions between the sectors, hence charge conserving operators must have a block-diagonal form with respect to the latter decomposition to charge sectors. 

The inclusion of an external gauge field term,  while promoting the symmetry to a local $U(1)$ gauge symmetry, leaves the essential features we discussed intact. However charge conservation gives rise to a meaningful notion of a local conservation of charge; 
for instance, the external gauge field may produce charge, but only pairs of opposite sign.

\section{ External gauge field}\label{sec:intHam}

As we have seen, the interaction with the external gauge field involves an electric (scalar) interaction \eqref{relIntHam}, and a magnetic (vector) interaction  \eqref{vectorrelIntHam}.

We next propose methods for realizing the above interactions. The realization of the electric part is more straightforward, and involves only additional laser-induced interaction coupling terms   between  the condensates. 
The magnetic  term $\mathcal{U}_B^{(1)}$ , involves spatial derivatives, and turns more challenging. We show in  \ref{subsec:vector},  that $\mathcal{U}_{b}^{(1)}$ can be obtained with the help of an ancillary system.

\subsection{Electric interactions}\label{subsec:scalar}
\begin{figure}[t]
                \includegraphics[width=0.2\textwidth]{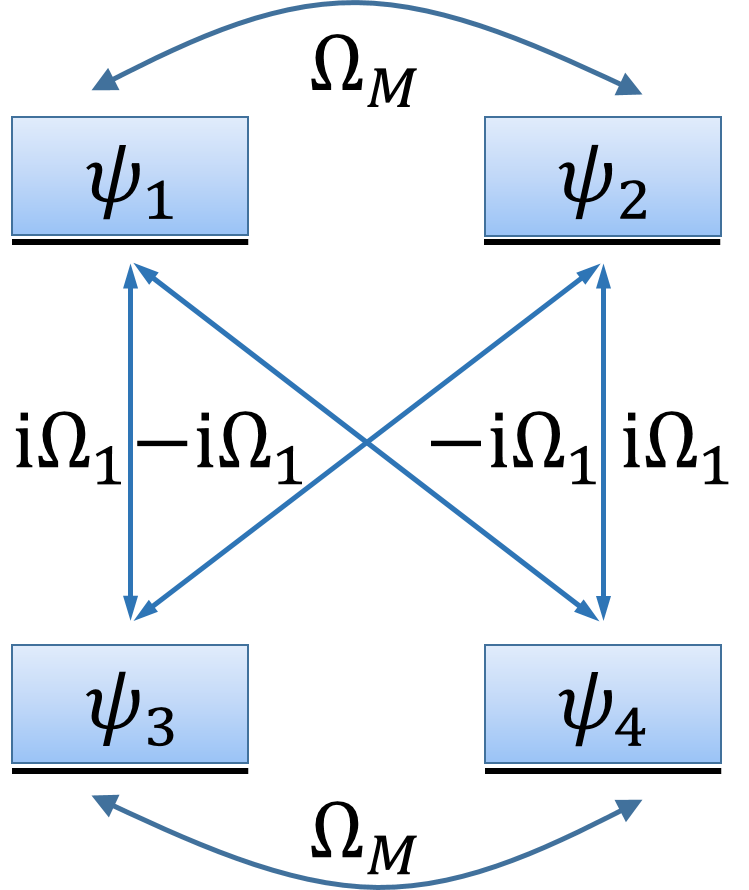}
                \caption{Interaction of the complex field and the electric potential $(\mathcal{U}_{E}^{(1)})$, in terms of the analog condensates and the laser's Rabi frequency}
                \label{fig:interaction1}
\end{figure}

The scalar potential interaction contains, to the first order in the charge $e$,  quadratic products of the field and its conjugate momentum,
\begin{equation}\label{scalarinte1}
\mathcal{U}_{E}^{(1)} = ieA^{0}\left(\Pi^{\dagger}\Phi^{\dagger}-\Phi\Pi\right)~.
\end{equation}
In terms  of the condensates' fields this gives
\begin{equation}
\mathcal{U}_{E}^{(1)}\propto \left(\psi_1^{\dagger}\psi_3-\psi_1^{\dagger}\psi_4-\psi_2^{\dagger}\psi_3+\psi_2^{\dagger}\psi_4\right)-\mbox{h.c.}
\end{equation}
Observing the last equation, we see that the interaction can be realized by adding Laser Raman couplings, depicted schematically in Fig. \ref{fig:interaction1}.
We can then deduce the relation
\begin{equation}\label{elec_ident}
\Omega_1= 2eA^{0}.
\end{equation}

\vskip .3cm

Moving-on to consider the second order electric interaction term, we have
\begin{equation}\label{scalarinte2}
\mathcal{U}_{E}^{(2)}=e^{2}A^0 A_{0}\Phi^{\dagger}\Phi=e^{2}A^0 A_{0} \frac{\phi_{1}^{2}+\phi_{2}^{2}}{2}~.
\end{equation}
This has a form of a ``mass term", as it is proportional to the square of the fields $\phi_i$. Since the interaction does not couple the components of the complex field, one can build it separately  for each field components.

To construct terms of the form $\phi_i^{2}$, let us examine first the effect of laser induced self-interaction of a field excitation $\varphi_i$ with itself:
\begin{equation}\label{effint2}
\Omega_2\varphi^{\dagger}\varphi=\Omega_2\left(\pi^{2}+\phi^{2}\right)~.
\end{equation}
Here $\Omega_2$ is the magnitude of the effective Raman frequency of the driving laser.
The interaction strength will be determined by the magnitude of the Rabi frequency $\Omega_2$, and the magnitudes of $\phi^{2}$ and $\pi^{2}$.
From equations \eqref{relfield_bog_amps} and \eqref{conmomenta_bog_amps}, we see that the latter depend, in turn, on the amplitudes $u(\p)\pm v(\p)$. 
Using Eq. \ref{prefactors}, the interaction strength for an excitation with momentum $p$ in the phonon regime can be approximated by
\begin{equation}
\left\langle\Omega_2\varphi^{\dagger}\varphi\right\rangle_p\approx\left|\Omega_2\right|\left(\frac{E}{mc_s^2}+\frac{mc_s^2}{E}\right)~.
\end{equation}
Restricting the Raman frequency $\Omega_2$ to be of the same order of magnitude as that of the phonons energy, the interaction's strength becomes
\begin{equation}
\left\langle\Omega_2\varphi^{\dagger}\varphi\right\rangle_p\approx\frac{E^{2}}{mc_s^{2}}\left(\frac{E}{mc_s^2}+\frac{mc_s^2}{E}\right)=E\left(\left(\frac{E}{mc_s^2}\right)^{2}+1\right)~.
\end{equation}
We observe that in the phonon regime, the first term does not contribute in the leading order, and can be neglected in the tree level Hamiltonian.
Namely, if the strength of the Raman frequency is in the order of the phonon's energy,  then $\varphi\approx-\varphi^{\dagger}$.

\begin{figure}[t]
                \includegraphics[width=0.2\textwidth]{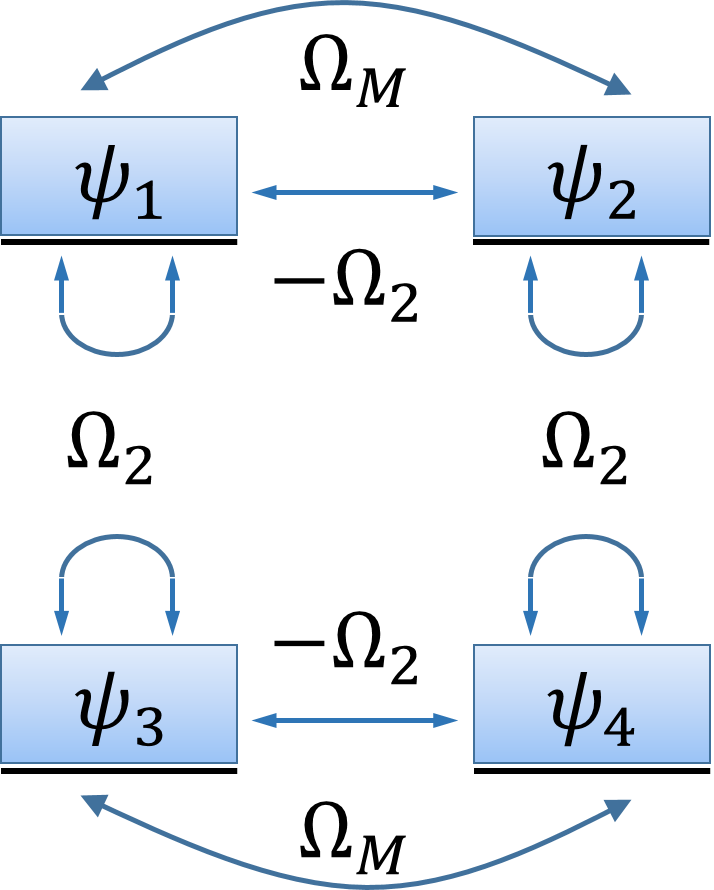}
                \caption{Interaction of the squared gauge field and the charged scalar field $(\mathcal{U}_{E}^{(2)}$ or $\mathcal{U}_{B}^{(2)})$, given in terms of the analog condensates and the laser's Rabi frequency}
                \label{fig:interaction2}
\end{figure}

We therefore conclude, that if the Raman frequency is bounded as described above, the self-interaction term \eqref{effint2} gives rise to 
\begin{equation}
2\Omega_2\varphi^{\dagger}\varphi\approx\Omega_2\phi^{2}.
\end{equation}
Using \rf{eigenmodes}, we find that the required interaction is
\begin{eqnarray}
\mathcal{U}_{\text{scalar}}^{(2)}=\Omega_2\left(\psi_1^{\dagger}\psi_1+\psi_2^{\dagger}\psi_2-\psi_2^{\dagger}\psi_1-\psi_1^{\dagger}\psi_2\right.\nonumber\\
\left.\psi_3^{\dagger}\psi_3+\psi_4^{\dagger}\psi_4-\psi_4^{\dagger}\psi_3-\psi_3^{\dagger}\psi_4\right)~.
\end{eqnarray}
We can now  identify the relation between the Raman frequency $\Omega_2$, and the square of the external electric potential $eA^{0}$: 
\begin{equation}\label{sqrd_elec_ident}
\Omega_2=\frac{\left(eA^{0}\right)^{2}}{mc_s^{2}}~.
\end{equation}
The construction of $\mathcal{U}_{\text{scalar}}^{(2)}$ in terms of the required Laser couplings, is depicted schematically in Fig. \ref{fig:interaction2}.

In order to match correctly the relative strength of the electric interaction, (\ref{scalarinte1},\ref{scalarinte2}), we need to tune the ratio of $\Omega_1$ and $\Omega_2$ such that
\begin{equation}
\frac{\Omega_{2}}{\Omega_{1}}=\frac{eA^{0}}{mc^{2}}~.
\end{equation}

\subsection{Magnetic interactions}\label{subsec:vector}
The interaction of the charged field with the vector potential \eqref{vectorrelIntHam} contains two terms.
The second order (``Landau") interaction term, has a similar structure as the second order (``Schwinger") electric term \eqref{scalarinte2}.
Therefore it can be also realized in the same manner.
We can, include the second order magnetic contribution simply by re-defining $\Omega_2$ as
\begin{equation}
\Omega_2=\frac{e^{2}A^{\mu}A_{\mu}}{mc_s^{2}}~.
\end{equation}
We note that $\Omega_2$ also controls the strength of the vector potential interaction, and thus the coupling strength of the laser that gives rise to $\mathcal{U}_{B}^{(1)}$ needs to be tuned as well.

The first term in \rf{vectorrelIntHam}, $\mathcal{U}_{B}^{(1)}$, involves spatial derivatives of the complex field. In terms of the complex field components  
\begin{eqnarray}\label{evecint}
\mathcal{U}_B^{(1)} &=& ieA^{i}\Phi^{\dagger}\cgrad_{i}\Phi\\
&=& ieA^{i}\left(\phi_1\cgrad_i\phi_1-i2\phi_1\cgrad_i\phi_2+\phi_2\cgrad_i\phi_2\right)~.\nonumber
\end{eqnarray}

To construct interaction of this sort, we now include supplementary 
ancillary condensates, which will be referred to as the ``virtual system", and use the subscript $V$ to denote them.
The condensates which describes the physical (``interacting") system, will be denoted by a subscript $I$.

The key idea will be then to use the ancillary condensate as intermediate ``virtual levels", and to obtain the derivative terms from second order Raman transitions.
Due to the continuous spectrum of the Bogoliubov Hamiltonian and the collective nature of its excitations, a virtual transition of atoms between the two systems may excite collective excitations in the virtual system.
This can also produce undesired correlations, and transfer energy between $I$ and $V$.

\begin{figure}[t]
                \includegraphics[width=0.2\textwidth]{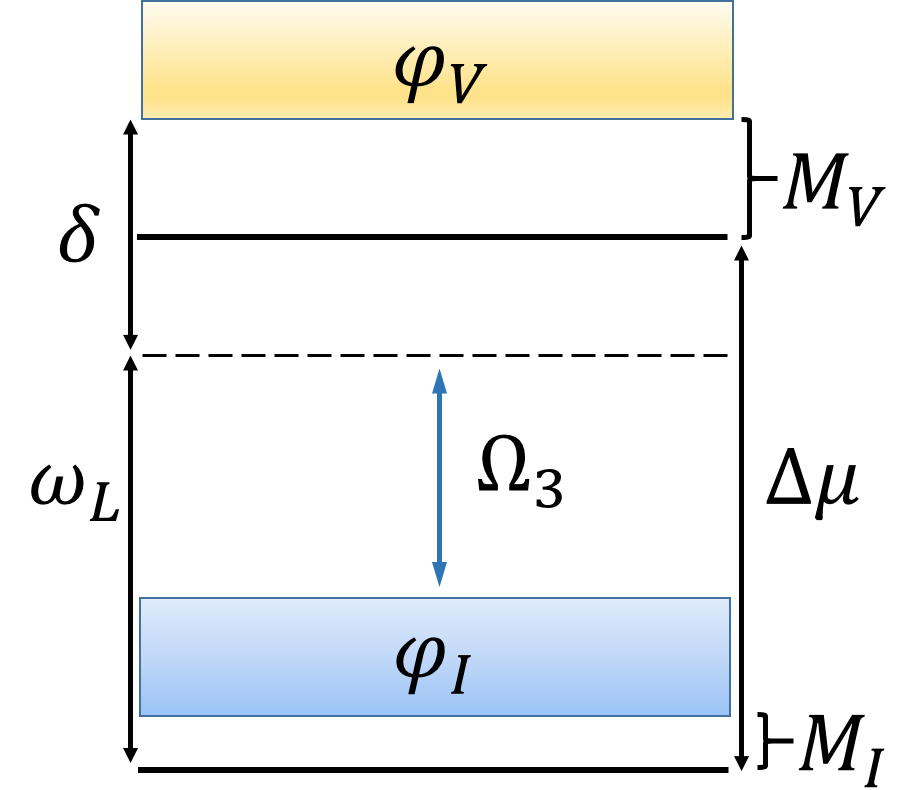}
                \caption{Interaction between the massive modes in the simulating (blue), and the ancillary (yellow) system.}
                \label{fig:toyinteraction}
\end{figure}

\noindent To avoid this, we shall introduce a gap in the virtual condensate, by using a massive eigenmode.
Therefore our ancillary system is built of two coupled condensates.
Only the massive mode of $V$ will be coupled to the interacting system.
To avoid the undesired correlations, we set the excitation's effective mass of the virtual condensate $M_V$, to be less but of the order of $m_I$.
Assuming that the excitations $\varphi_{I}$ are in the long wavelength regime ($p_I\ll m_Ic_{sI}$), their energy scale is then much smaller than the one of $\varphi_{M_V}$.
As a result, transitions of atoms between the virtual and interacting systems cannot produce excitations in the ancillary system due to energy conservation.
Furthermore we wish to prevent excitations in the interacting system due to transitions from the virtual one.
In order to do so, we should reduce the temperature of the virtual system below the effective rest energy of the massive excitations, i.e.
\begin{equation}
k_{B}T\ll M_{V}c_{sV}^{2}~.
\end{equation}
These energy constraints produce an additional energy gap between the interacting and virtual subsystems (as depicted in Fig. \ref{fig:toyinteraction}), where the difference in the two speeds of sounds yields a gap between the two chemical potentials.
 
Let us now construct a self interaction of a massive excitation that contain first order derivative.
Having discussed the properties of the ancillary system $V$, let us proceed to study the effect of (virtual) second order transitions, that arise from the laser induced interaction between the virtual and interacting systems:
\begin{equation}
H_{\text{int}}=\int d\x \Omega_3(\x)\left(\varphi_{V}^\dagger(\x)\varphi_{I}(\x)+\text{h.c.}\right)~,
\end{equation}
depicted in Fig. \ref{fig:toyinteraction}.

Using the method of adiabatic elimination \cite{cohen1992atom}, we expand the interaction to second order in the small parameter $\Omega_{3}/\delta^{2}$. The resulting effective self interaction of $\varphi_{I}$, is then given by:
\begin{equation}\label{primvecint}
H_{\text{int}}^{(2)}=\!\!\int \!\! d\x d\y \Omega^{*}(\x)\Omega(\y)\varphi_{I}^\dagger(\x)\varphi_{I}(\y)\mathcal{I}(\x-\y)~.
\end{equation}
Here the function $\mathcal{I}(\x-\y)$ is the result of integrating-out the intermediate virtual states, 
\begin{eqnarray}
\mathcal{I}(\x-\y)&=&\frac{\left\langle\varphi_{V}^{\dagger}(\x)\varphi_{V}(\y)\right\rangle}{\delta} \nonumber\\
&=&\int d\p\frac{|v(\p)|^{2}}{\delta}e^{-i\p(\x-\y)}~,
\end{eqnarray}
where $\delta=\Delta\mu+\omega_{M_V}-(\omega_L+\omega_{M_I})\equiv \Delta\mu+\omega_{M_V}-\Delta$.
The function $\mathcal{I}(\x-\y)$ is numerically evaluated in appendix \ref{app:vecint}.

\vskip .3cm
Revising the energetic constraints on the interacting and the virtual systems, yield further relations between the length scale of the interacting and virtual systems.
Since $m_Ic_{sI}^{2}\lesssim M_Vc_{sV}^{2}$, and $\x-\y\sim (M_Vc_{sV})^{-1}$ (as is shown in appendix \ref{app:vecint}), the interaction length scale is comparable to the healing length of the interacting condensate, $\x-\y \lesssim (m_Ic_{sI})^{-1}$.
Thus in terms of a new set of coordinates
\begin{equation}
\begin{cases}
\x=\frac{1}{2}\left(\xmin+\xplus\right)\\
\y=\frac{1}{2}\left(\xmin+\xplus\right)
\end{cases}~,
\end{equation}
the effective interaction \eqref{primvecint} can be expanded in terms of $\xmin$ around $\xplus$.
Since \eqref{primvecint} is rotational symmetric, this expansion contains only derivatives of even order, thus the first order derivative terms which we seek are excluded.
Nevertheless, giving the Rabi frequency a harmonic profile, $\Omega_3(\x)\rightarrow \exp\lbrace i\kk\x\rbrace\Omega_3(\x)$, breaks the rotational symmetry, and defines the derivative's orientation.
\begin{figure}[t]
                \includegraphics[width=0.2\textwidth]{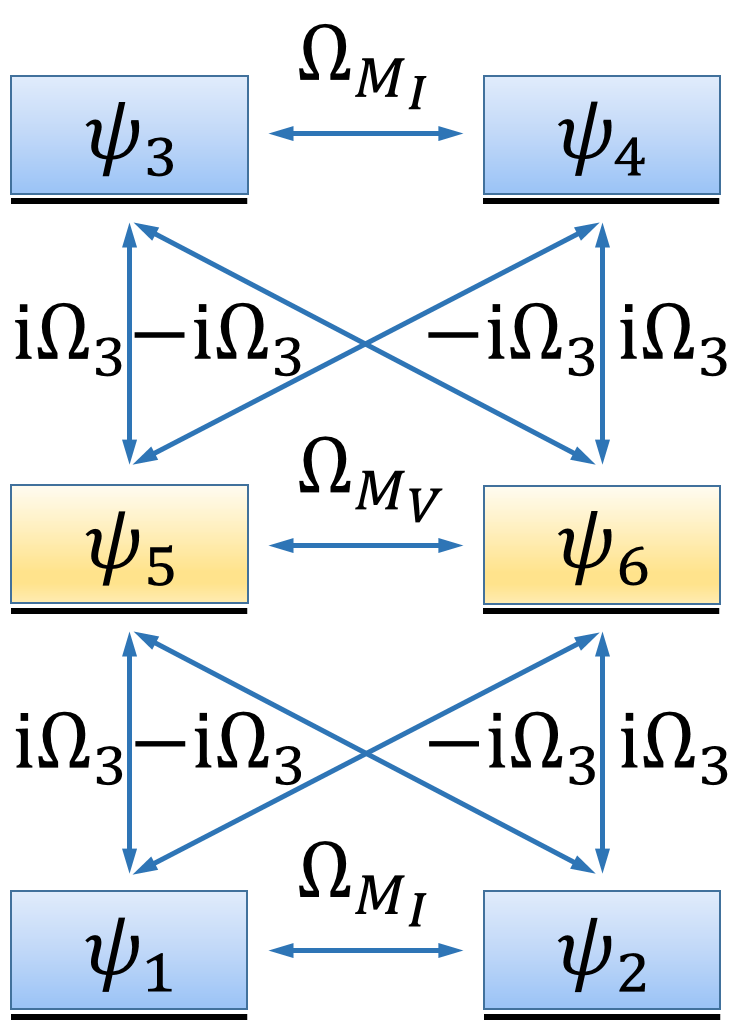}
                \caption{The full interaction between the different condensates in the simulating (blue) and ancillary (yellow) systems, given in terms of the analog condensates and the driving laser's Rabi frequency. This construction leads to the effective interaction, which in turn yields the interaction of the charged field with the vector potential}
                \label{fig:vecint}
\end{figure}
\noindent Expanding the effective interaction to first order in $\xmin$ yields two terms.
The first $(\propto \xmin^{0})$ has the form of a second order Raman transition, and thus can be compensated (see appendix \ref{app:vecint}).
The second term $(\propto \xmin)$ contains first order derivatives oriented in the direction of $\kk$, and is given by
\begin{equation}\label{effvecint}
H_{\text{eff}}=i\!\!\int\!\! d\xplus\Omega_3^2(\xplus)\mathcal{G}\varphi_{I}^\dagger(\vxplus)\cgrad_{\kk}\varphi_{I}(\vxplus)~,
\end{equation}
where $\mathcal{G}=\mathcal{G}(\kk,m,c_{s_I},c_{s_V})$ is a real function (see appendix \ref{app:vecint}).

Substituting the massive excitation in terms of the scalar field and its conjugate momentum (\ref{map_relfield},\ref{map_conmomenta}) in the latter, leads to terms that breaks the local U(1) invariance of the theory.
In order to avoid them, the effective Rabi frequency $(|\Omega_3|^2\mathcal{G})$ should be bounded, following similar arguments used in the construction of \eqref{scalarinte2}.
Assuming that the energy scale of the massive excitations is much smaller than the cutoff energy, and that $|\Omega_3|^2\mathcal{G}\lesssim Mc_{sI}^{2}$, we obtain
\begin{equation}
2\varphi_{I}^\dagger\cgrad_{\kk}\varphi_{I}\approx\phi\cgrad_{\kk}\phi~.
\end{equation}

\begin{figure}[t]
                \includegraphics[width=0.27\textwidth]{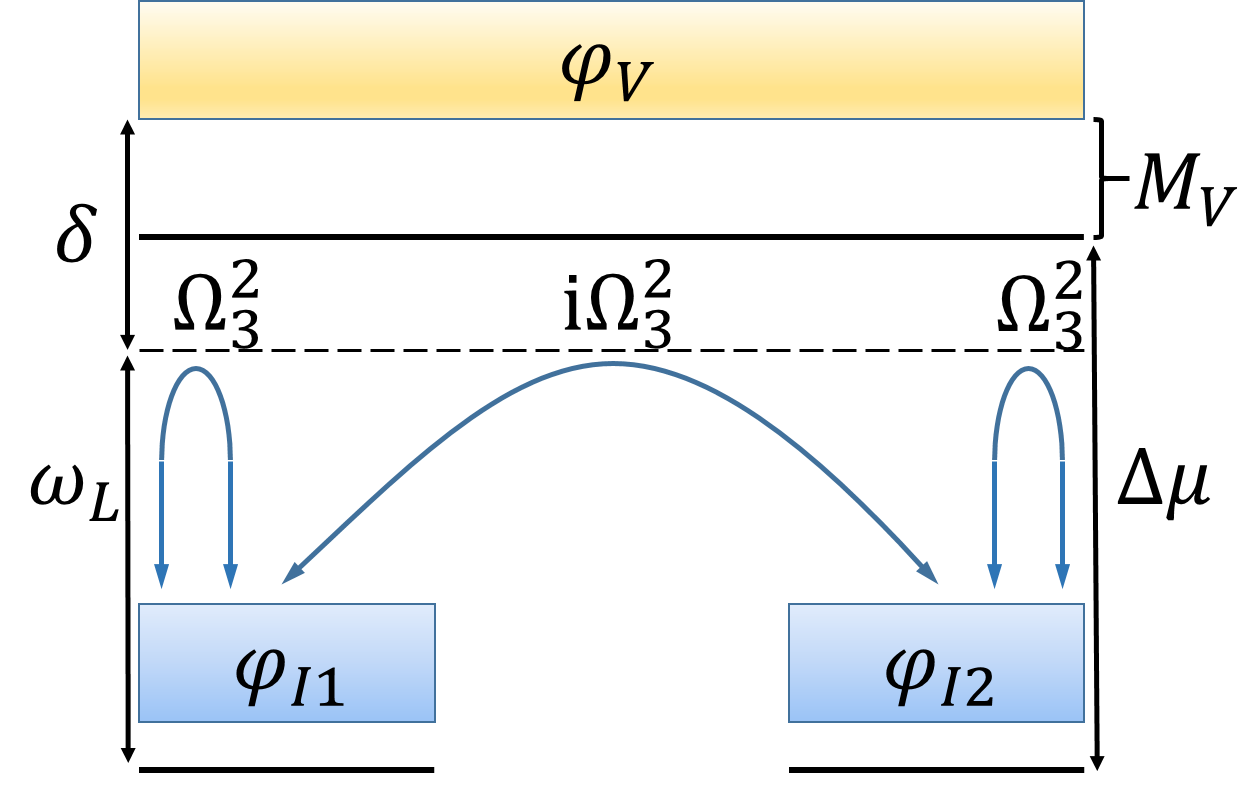}
                \caption{The effective interaction between the modes in the interacting system}
                \label{fig:effvecint}
\end{figure}

Finally, in order to construct the full interaction \eqref{evecint}, we need:
\begin{equation}
H_{\text{int}}=\int\!\! d\x~\Omega(\x)(\varphi_{I1}^\dagger(\x)+i\varphi_{I2}^\dagger(\x))\varphi_{V}(\x)+\mbox{h.c.}~.
\end{equation}
In Fig. \ref{fig:vecint}, we describe the required interactions between BEC components. The resulting effective interaction is schematically depicted in Fig.\ref{fig:effvecint}.

Finally, we note that the resulting effective Rabi frequency relates to the external vector potential according to:
\begin{equation}
eA^{i}=\Omega_{i}^{\text{eff}}\equiv\Omega_{3i}^{2}(\x)\,\mathcal{G}(mc_{sV},k_i)mc_{sI}~.
\end{equation}

\section{Discussion}
In this article, we proposed a method for simulating a QFT system of a charged scalar field coupled
to an external electromagnetic field.
Unlike previous studied analogies of fields in curved space-time , which require accelerating the atomic condensate, in the present method the effect of the electromagnetic field is generated in a static condensate. The interaction here is introduced by means of external lasers that couple the internal atomic levels in a certain manner.
Therefore we expect that such a system will be relatively more stable - there are no special requirements nor special limitations on the mean field properties.
Furthermore, this provides a wider flexible range of applicability in controlling the various electromagnetic external fields that one may wish to apply on the ``charged" phonons.
In particular, it may enables the study of the Sauter potential and its impact on the
charged field, and especially in its two asymptotic limits
which yields the the Klein paradox and the Schwinger
effect. In addition to the scattering properties which can
be experimentally measured in this sort of simulations, it may be even more intriguing to observe and study the process of pair creation, and vacuum instability due to an electric field. Moreover, electric potential of this type may also be employed as a source of particles or anti-particles for other quantum simulations which involves bulk excitations, such as gravitational analogies and QFT in curved space-time.

\bibliography{refs}

\appendix

\section{Diagonalization of the Bogoliubov Hamiltonian for two mode BEC}\label{app:diagonalization}
In order to find the full dispersion relation and normal modes of our two mode BEC \eqref{GPhamiltonian}, we shall derive the Bogoliubov Hamiltonian.
Substitution of the mean field expansion into the GP Hamiltonian \eqref{GPhamiltonian}, yields a Hamiltonian of the form $H_0+H_p$, where $H_0$ is the ground state energy (zero momentum), and $H_p$ is the quadratic part that describes the excitations from the condensate.
Next, in order to constrain conservation of particles number, we plugging
\begin{equation}
N_{0}=N_i-\sum_{p\neq 0}^{}\cop{a}{i,p}\aop{a}{i,p}
\end{equation}
into $H_0$, keeping only quadratic terms of the form $\cop{a}{i,p}\aop{a}{i,p}$ \cite{pitaevskii2003}.
Equivalently we can use the operator $\hat{K}=\hat{H}-\mu\hat{N}$, where the last term acts as a Lagrange multiplier \cite{Pethick2002}.
The resulting Bogoliubov Hamiltonian is
\begin{widetext}
\begin{eqnarray}\label{ham}
H_p=\sum_{j,\p \neq 0}\left(\frac{p^2}{2m}+nU-\Omega\right)\cop{a}{j,\p}\aop{a}{j,\p}+\frac{1}{2}nU(\cop{a}{j,\p}\cop{a}{j,-\p}+\aop{a}{j,\p}\aop{a}{j,-\p})\nonumber\\
+nU^{\prime}(\cop{a}{2,\p}\cop{a}{1,-\p}+\aop{a}{1,\p}\aop{a}{2,-\p})+(nU^{\prime}+\Omega)(\cop{a}{2,\p}\aop{a}{1,\p}+\cop{a}{1,\p}\aop{a}{2,\p})~.
\end{eqnarray}
\end{widetext}
Since $H_{p}$ is quadratic in terms of the field operators $\aop{a}{j,\p}$ and $\cop{a}{j,\p}$, it can be diagonalized by a linear transformation, leading to a new set of field operators.
In order to find the appropriate transformation, we adopt the diagonalization method in \cite{diagonalization1}.
We begin by introducing the vector
\begin{equation}\label{avec}
\eta^\dagger=(\cop{a}{1,\p},\cop{a}{2,\p},\aop{a}{1,-\p},\aop{a}{2,-\p})~.
\end{equation}
$H_{p}$ can now be written now as
\begin{equation}\label{nondiag}
H_p=\eta^\dagger\mathcal{H}\eta~.
\end{equation}
We then define the transformation 
\begin{equation}\label{lintrans}
\xi^\dagger=\eta^\dagger T \,\,;\,\, \xi=T^\dagger\eta~,
\end{equation}
where $T$ is a $4\times 4$ matrix, and $\xi$ is composed by the ``free'' field operators $\aop{b}{j,\p}$~.
The bosonic commutation relation of the new field operators, $\aop{b}{i,\p}$, implies a constraint on the transformation matrix $T$, by which the inverse of $T$ is defined.
Finally, plugging the transformation \eqref{lintrans} into the Hamiltonian \eqref{nondiag}, yields the eigenvalue equation
\begin{equation}\label{eveq}
T^{-1}\mathcal{H}JT=\mathcal{H}_DJ=
\begin{pmatrix}
	E & 0 \\
	0 & -E
\end{pmatrix},
\end{equation}
where $E$ is a $2\times 2$ diagonal matrix of the eigenenergies $E_j$.
The matrix $J=\mbox{diag}\{I_N,-I_N\}$ is closely related to the symplectic matrix, which is widely used in the diagonalization process of a quadratic Hamiltonian given in canonical coordinates $\{p_n,q_n\}$.

The resulted dispersion relations of the two eigenmodes are
\begin{eqnarray}\label{phen}
E_1^2&=&2\left(nU+nU^{\prime}\right) \epsilon _p+\epsilon _p^2\\
&=& 2mc_{s1}^{2}\epsilon_p+\epsilon _p^2~,\nonumber\\
E_2^2 &=& (nU-nU^{\prime}-2\Omega)^2-(nU-nU^{\prime})^2\\
&&+2(nU-nU^{\prime}-2\Omega) \epsilon _p\!+\!\epsilon_p^2\nonumber\\
&=&E_\rr^2+2mc_{s2}^{2}\epsilon_p+\epsilon _p^2~\nonumber.
\end{eqnarray}
As one can see, the energies differs from the well known phonon energy of a one component BEC.
In this set-up one of the modes acquires an effective mass; a momentum free term, which depends only on the coupling constants of the system.
In addition, the speed of sound which can be identified as the factor of the ``kinetic energy" ($\propto \epsilon_p$), is different for every mode.
Hence the massive and massless eigenmodes, exhibiting two different sound cones and thus have different causal structure.

On the other hand, the elements of the transformation matrix $T^{-1}$, have the generic form \cite{pitaevskii2003}
\begin{equation}
u_{J}(\p),v_{J}(-\p)\propto\pm\sqrt{\frac{ \left(\text{mc}_{sJ}^2+\epsilon _p\right)}{2E_J}\pm \frac{1}{2}}~,
\end{equation}
where $J$ refers to the two eigenmodes above.
\begin{figure}[t]
		\includegraphics[width=0.5\textwidth]{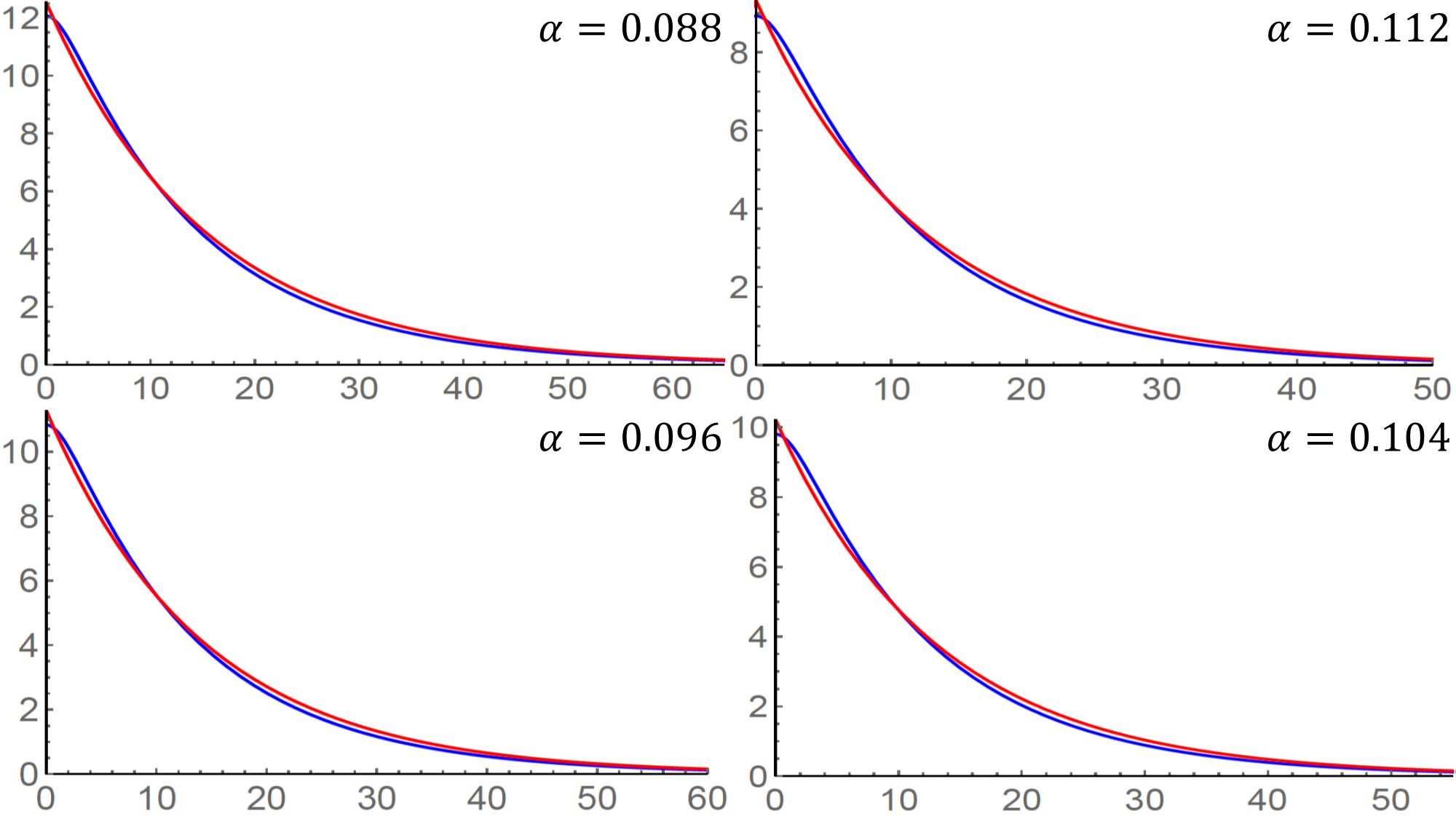}
                \caption{1D. A comparison between the numerical result (blue line), and the evaluated function (red line) for different values of $\alpha$.}
                \label{fig:comparison}
\end{figure}
\section{The effective interaction}\label{app:vecint}
In order to derive Eq. \ref{effvecint}, we begin with the evaluation of $\mathcal{I}(\xmin)$.
We shall find an analytic function that optimally describes the behavior of $\mathcal{I}(\xmin)$.
Then, we will use the resulting function to find $\mathcal{G}$, and obtain the effective Rabi frequency $\Omega_{i}^{\text{eff}}$.

\vskip .3cm

The function $\mathcal{I}(\xmin)$ is given by
\begin{eqnarray}
\mathcal{I}(\xmin)&=&\frac{\left\langle\varphi_{V}^{\dagger}(\x)\varphi_{V}(\y)\right\rangle}{\delta} \nonumber\\
&=&\int d\p\frac{|v(\p)|^{2}}{\delta}e^{-i\p\vxmin}~,
\end{eqnarray}
where $\delta=\Delta\mu+\omega_{M_V}-(\omega_L+\omega_{M_I})\equiv \Delta\mu+\omega_{M_V}-\Delta$.

In order to evaluate the optimal function we shall proceed as follows:
First, we define the ratio of the effective and the atomic masses, $\alpha=M_V/m_V$.
Since the interaction is off resonance, the detuning of the laser, and the chemical potential can be determined by this ratio. Thus we are left with a controlled parameter, $\alpha$, which is restricted to be much smaller than one.
Next, we choose a function, and fit it to the numeric solutions for various values of $\alpha$.
Then, we find the dependence of the analytic function's parameters on $\alpha$.

\noindent In the proceeding calculation we set $c_s=1$, and fix the parameters as follows:
\begin{equation}\label{paramDef}
\begin{cases}
0.08\leq \alpha\leq 0.12\\
\Delta\mu-\Delta=M
\end{cases}
\end{equation}
\begin{figure}[t]
		\includegraphics[width=0.5\textwidth]{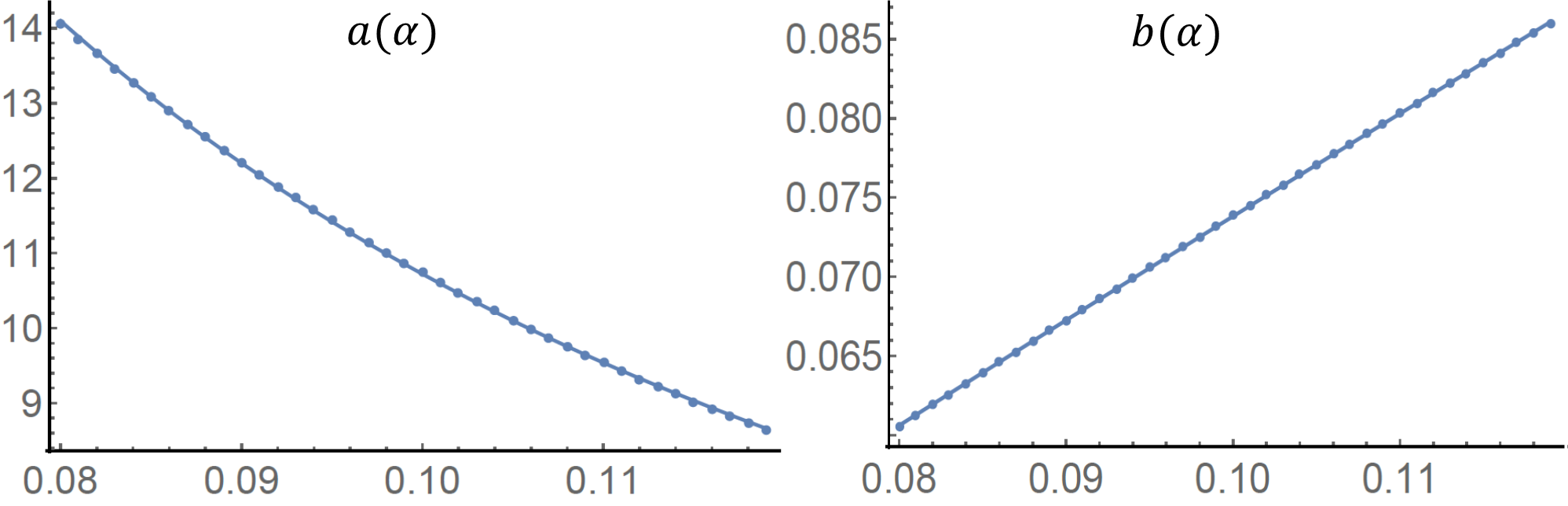}
                \caption{1D. A comparison between the numerical results of the parameters $a$ and $b$ for different values of $\alpha$ (dots), and their evaluated functions (solid line). The functions $a(\alpha)$ and $b(\alpha)$ are given in (\ref{a1d},\ref{b1d}).}
                \label{fig:parameters behaviours}
\end{figure}

In 1D, the function $\mathcal{I}(\xmin)$ is given by
\begin{eqnarray}
\mathcal{I}(\xmin)&=&\int_{-\infty}^\infty\!\!dpe^{ip\xmin}\frac{\left(\frac{p^2}{2 m}+m-E(p)\right)}{E(p) \left(\Delta\mu+E(p)-\Delta \right)}\\
&&=\sqrt{2}\int_{-\infty}^\infty\!\!d\eta e^{i\eta s}\frac{\left(1+\eta^2-E(\eta)\right)}{E(\eta) \left(\alpha+E(\eta)\right)} ~,\nonumber
\end{eqnarray}
where $s=\sqrt{2}m_V\xmin$, and
\begin{equation}
E(\eta)=\sqrt{\alpha^2+2\eta^2+\eta^4}~.
\end{equation}
We choose $\mathcal{I}(s)$ to be
\begin{equation}\label{fitFunc}
\mathcal{I}(s)=a e^{-bs}~,
\end{equation}
where $a$ and $b$ are both functions of $\alpha$.
A comparison between the Numerical result and the fitted function is given in Fig. \ref{fig:comparison} for different values of $\alpha$.
We observe that the chosen function fits well to the numerical result, and describing the general behavior of the numeric function.
Next, we shall find the dependence of the two parameters $a$ and $b$ on $\alpha$.
In order to do so, we take the results of the two parameters for various values of $\alpha$, and fit a function that describes their dependency on the parameter $\alpha$. The two parameters of the approximated function resulting in
\begin{eqnarray}
a(\alpha)&\approx&\frac{0.64}{\alpha^{1.23}},\label{a1d}\\
b(\alpha)&\approx&0.56 \alpha^{0.88}~,\label{b1d}
\end{eqnarray}
where a comparison of the numerical results and the fitted curves is given in Fig. \ref{fig:parameters behaviours}.
In conclusion, we obtain:
\begin{equation}
\mathcal{I}(\xmin)\approx \frac{0.9}{\alpha^{1.23}} e^{-0.8 \alpha^{0.88} m_V|\xmin|}~.
\end{equation}
As one can see, the decay rate of the function $\mathcal{I}(\xmin)$ is greater than $M_V$.
Hence it dictates a characteristic length scale, that is in the scale or smaller from the interacting modes healing length, $\xi_{I}\equiv (\sqrt{2}mc_{sI})^{-1}$.
\begin{figure}[t]
		\includegraphics[width=0.5\textwidth]{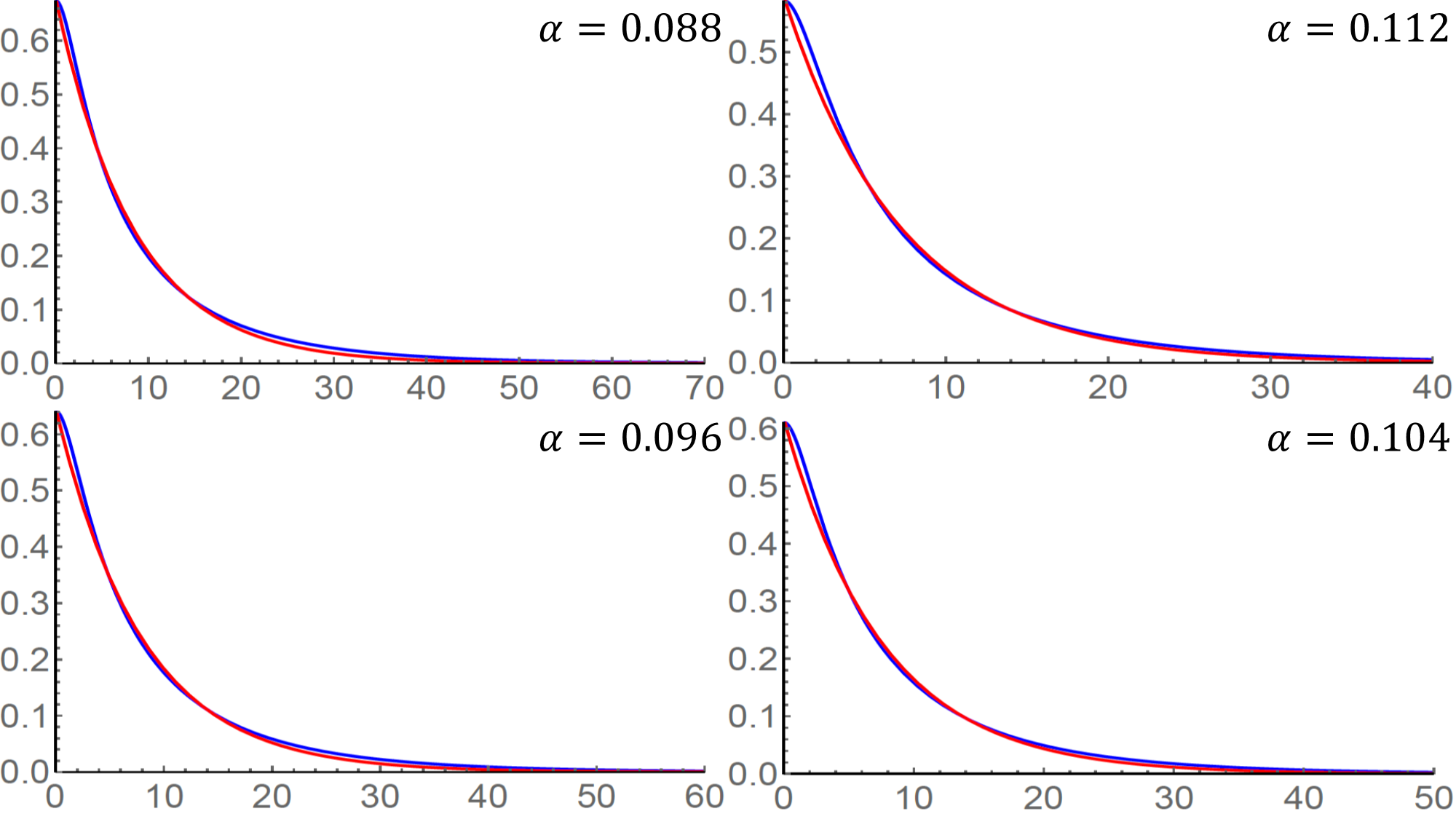}
                \caption{2D. A comparison between the numerical result (blue line), and the evaluated function (red line) for different values of $\alpha$.}
                \label{fig:comparison 2D}
\end{figure}
\vskip .3cm
In 2D, the function $\mathcal{I}(\xmin)$ is given by
\begin{equation}
\mathcal{I}(s)=4\pi m_V\int_0^\infty\!\!d\eta \frac{\left(1+\eta^2-E(\eta)\right)}{E(\eta) \left(\alpha+E(\eta)\right)}J_0(\eta s)~,
\end{equation}
where $J$ is a Bessel function of the first kind.

\begin{figure}[t]
		\includegraphics[width=0.5\textwidth]{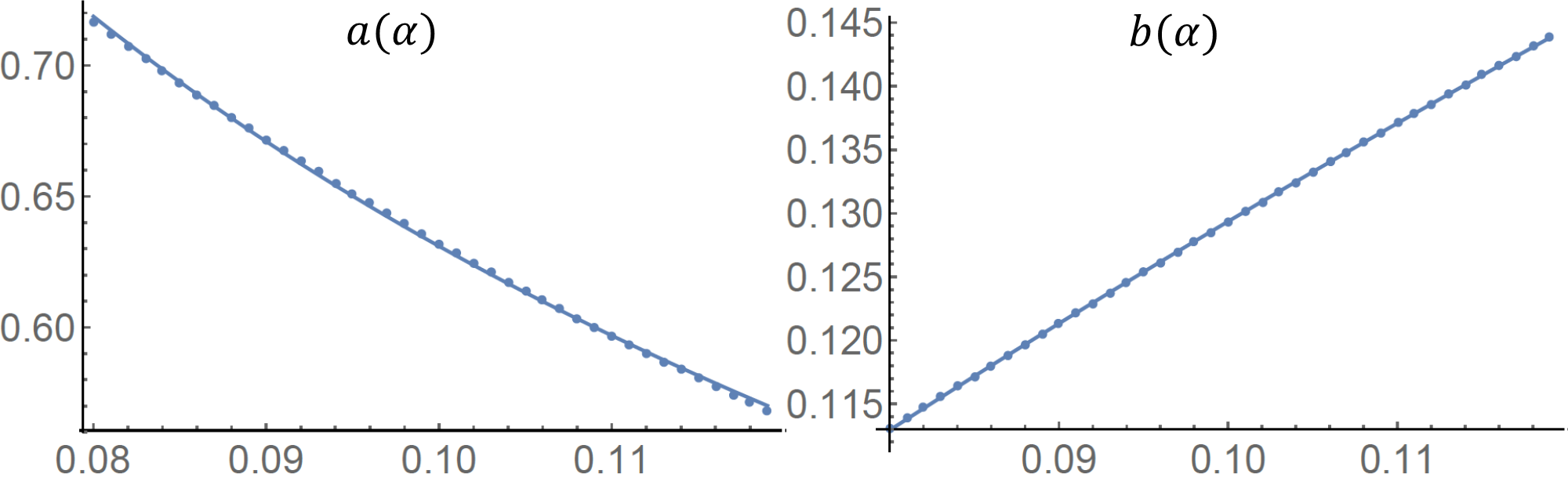}
                \caption{2D. A comparison between the numerical results of the parameters $a$  and $b$ \eqref{b2d}, for different values of $\alpha$ (dots), and their evaluated functions (solid line). The functions $a(\alpha)$ and $b(\alpha)$ are given in (\ref{a2d},\ref{b2d}).}
                \label{fig:parameters behaviours 2D}
\end{figure}

\noindent We repeat the procedure used in the case of one dimensional system, and choose the evaluated function to be (\ref{fitFunc}).
The comparison between the numeric results and the fitted function is depicted in Fig. \ref{fig:comparison 2D}.
The dependence of the parameters on the effective mass is given by
\begin{eqnarray}
a(\alpha)&=&\frac{0.16}{\alpha^{0.58}}~,\label{a2d}\\
b(\alpha)&=&0.523 \alpha^{0.61}~,\label{b2d}
\end{eqnarray}
and depicted in Fig. \ref{fig:parameters behaviours 2D}.
In conclusion, $\mathcal{I}(\xmin)$ can be evaluated by:
\begin{equation}\label{Isolution2d}
\mathcal{I}(\xmin)\approx \frac{2m_V}{\alpha^{0.58}}e^{-0.74\alpha^{0.61}m_V\xmin}~.
\end{equation}
We observe that the evaluated function decays in the scale or faster than $M_V^{-1}$, and hence the characteristic length scale of the interaction is in the order or smaller than the healing length of the interacting system.
In conclusion, the estimation of the interaction length scale $\xmin$ is justified, and the effective interaction \eqref{primvecint} can be expanded in terms of this relatively small parameter.

\vskip .3cm

We proceed with the formulation of the effective interaction \eqref{effvecint}.
Expanding \eqref{primvecint} to first order in $\xmin$ yields
\begin{equation}\label{nonCompInt}
H_{\text{eff}}=\int \!\! d\vxplus \Omega^2 (\xplus)\varphi_{I}^\dagger(\vxplus)\hat{\mathcal{O}}\varphi_{I}(\vxplus)~,
\end{equation}
where
\begin{equation}
\hat{\mathcal{O}}=\int d\xmin \mathcal{I}(\xmin)e^{i\kk \vxmin}\left(1+\vxmin\cdot\cgrad\right)~.
\end{equation}
Inserting \eqref{Isolution2d} into the latter, we can carry out the integration on $\xmin$, resulting in
\begin{equation}
\hat{\mathcal{O}}=\mathcal{F}+i\mathcal{G}\cgrad_{\kk}~,
\end{equation}
where $\mathcal{F}$ and $\mathcal{G}$ are given by
\begin{eqnarray}
\mathcal{F}&=&2m_Va\int_0^{\infty} d\xmin \xmin e^{-\sqrt{2}bm_V\xmin}J_0(k\xmin)\nonumber\\
&=&\frac{2\pi}{m_V}\frac{ab}{(b^{2}+\eta_k^2)^{3/2}}~,\\
\mathcal{G}&=&2m_Vb\int_0^{\infty}d\xmin \xmin^2 J_1(k \xmin) e^{-\sqrt{2}bm_V \xmin}\nonumber\\
&=&\frac{2\pi}{\sqrt{2}m_V^2}\frac{ab\eta_k}{(b^{2}+\eta_k^2)^{5/2}}~.
\end{eqnarray}
Here $\eta_k=k/\sqrt{2}m_V$, and the numerical coefficients $a$ and $b$ are given in \eqref{a2d} and \eqref{b2d}.
Finally, the effective interaction \eqref{primvecint} is
\begin{equation}
\mathcal{H}_{\text{eff}}=\Omega^2(\xplus)\varphi_{I}^\dagger(\vxplus)\left[ \mathcal{F}+i\mathcal{G}\cgrad_{\kk}\right]\varphi_{I}(\vxplus)~.
\end{equation}
The first term is irrelevant to our model, and thus should be compensated by an additional laser induced interaction,
\begin{equation}
\mathcal{H}_{\text{comp}}=\Omega_{\text{comp}}\left(\psi_1^{\dagger}\psi_1+\psi_2^{\dagger}\psi_2-\left(\psi_1^{\dagger}\psi_2+\psi_2^{\dagger}\psi_1\right)\right)
\end{equation}
where $\Omega_{\text{comp}}=-\Omega^2(\xplus)\mathcal{F}$~.

\end{document}